\documentclass[12pt, draftclsnofoot, onecolumn]{IEEEtran}
\ifCLASSINFOpdf
\else
   \usepackage[dvips]{graphicx}
\fi
\usepackage{url}

\usepackage[english]{babel}
\usepackage[utf8]{inputenc}
\usepackage{graphicx}
\usepackage{amsmath}
\usepackage{amsfonts}
\usepackage{amssymb}
\usepackage{algorithm}
\usepackage{algorithmic}
\usepackage{float}
\usepackage{caption,subcaption,xspace,xcolor,cite}
\usepackage[normalem]{ulem}

\begin{document}

\title{Channel Estimation for Flexible Intelligent Metasurface Aided MIMO Communications}

\author{Vinícius L. Romano, André L. F. de Almeida, Daniel C. Araújo}

\maketitle

\begin{abstract}
Flexible Intelligent Metasurfaces (FIMs) enable wireless systems to adapt their three-dimensional geometry through morphing, thereby providing new spatial degrees of freedom. However, continuous deformation complicates the accurate acquisition of Channel State Information (CSI). This work proposes a multidimensional framework for MIMO systems with active FIM arrays at both the transmitter and receiver. A split single-time-scale training protocol sequentially introduces spatial variation by morphing the receiver, then the transmitter. The resulting signal model is formulated as a PARAFAC decomposition, and an alternating least squares (ALS) algorithm is employed to estimate steering matrices and path gains. Our numerical results show that the proposed channel estimation method yields accurate CSI recovery for different system setups.
\end{abstract}


\section{Introduction}

Reconfigurable intelligent surfaces (RISs) have emerged as a promising technology for wireless communications because they enable the propagation environment itself to be engineered rather than simply adapted to. By controlling the electromagnetic response of many low-cost elements, RIS-assisted systems can shape the amplitude and phase of the impinging wavefront and, consequently, reduce part of the randomness inherent to wireless propagation. This potential has motivated a broad literature on metasurface-assisted communications, including channel modeling, signal processing, and transceiver design problems. Nevertheless, most available architectures still rely on rigid and nearly planar surfaces, which limit their adaptability in dynamic environments.

Among the main challenges in RIS-assisted communications, channel estimation remains one of the most critical because accurate channel state information (CSI) is essential for configuring the surface response and optimizing data transmission. A large body of work has studied this problem from different perspectives, including tensor-based modeling, structured and unstructured channel estimation, and semi-blind processing strategies ~\cite{Almeida2007,confac,FAVIER_2012,Favier_EUSIPICO, deAraujoSAM2020,de2021channel,deAlmeida_2025_BDRIS,Nwalozie_2025,Tapie_2026}. In conventional RIS architectures, however, the surface geometry is typically fixed, so the estimation task primarily concerns the electromagnetic response of a static array.

In this context, Flexible Intelligent Metasurfaces (FIMs) extend the RIS concept by introducing mechanical reconfiguration in addition to electromagnetic control~\cite{An_2025_FIM,An_2025_FIM_MISO,Yang_2026_FIM_Architecture,An_2024_FIM_GLOBECOM,An_2025_Intelligent_Metasurfaces_Survey}. Unlike rigid metasurfaces, FIMs are composed of soft arrays whose radiating elements can independently change their physical positions, typically along the surface normal direction. This morphing capability enables the surface to adapt its three-dimensional (3D) shape to the propagation conditions, creating new spatial degrees of freedom and potentially improving coverage, diversity, and spectral efficiency. On the other hand, the same flexibility makes CSI acquisition more difficult, since the channel now depends not only on the propagation paths but also on the instantaneous geometric configuration of the transmitting and receiving surfaces.

For dual-FIM MIMO systems, where both the transmitter (Tx) and receiver (Rx) can independently morph, channel estimation becomes a highly coupled multidimensional problem. Recent works have begun to address this challenge by explicitly modeling the dependence between the FIM shape and the resulting channel. In~\cite{Xiao_2026}, the authors tackle the difficulty of mapping the channel under the virtually continuous set of possible FIM deformations using a learning-based hierarchical Fourier neural operator (H-FNO) that captures the nonlinear relation between morphology and channel response, reducing the number of required measurements with respect to classical strategies. In turn,~\cite{Jiang_2025} actively designs the FIM deformation to improve the sensing conditions of sparse recovery methods such as orthogonal matching pursuit (OMP), yielding more accurate estimates of path directions and gains, and subsequently a higher downlink signal-to-noise ratio. However, neither~\cite{Xiao_2026} nor~\cite{Jiang_2025} addresses the joint channel estimation problem when both ends are equipped with morphing FIM arrays, where the channel depends on two independent morphing configurations and existing single-FIM strategies become insufficient. To the best of our knowledge, this paper is the first to propose a tensor-based channel estimation framework for point-to-point MIMO systems with active FIM arrays at both ends. Specifically, the main contributions are:
\begin{itemize}
    \item We design a \emph{Split Single Time-Scale} training protocol that sequentially introduces spatial variation by morphing the Rx FIM while keeping the Tx FIM static, and vice versa. This strategy effectively decouples the highly correlated spatial dimensions of the channel.
    \item We develop a two-phase Alternating Least Squares (ALS) algorithm to jointly and iteratively estimate the morphed receive and transmit steering matrices, as well as the multipath gains, achieving robust CSI recovery.
\end{itemize}

\textit{Notation:} Scalars are denoted by lower-case letters ($a, b, \dots$), vectors by lower-case boldface letters ($\mathbf{a}, \mathbf{b}, \dots$), matrices by upper-case boldface letters ($\mathbf{A}, \mathbf{B}, \dots$), and tensors by calligraphic letters ($\mathcal{A}, \mathcal{B}, \dots$). The transpose and conjugate transpose are denoted by $(\cdot)^T$ and $(\cdot)^H$, respectively. The Khatri--Rao and Hadamard products are denoted by $\diamond$ and $\odot$, respectively. The Frobenius norm is denoted by $\|\cdot\|_{\mathrm{F}}$, $\sqcup_n$ denotes the concatenation operator along the $n$-th mode, and $\mathrm{diag}(\cdot)$ forms a diagonal matrix from a vector.

\section{System Model}

\subsection{System Configuration}

Let us consider a point-to-point MIMO communication system operating in a multipath environment, where the transmitter (Tx) and receiver (Rx) are Flexible Intelligent Metasurfaces (FIMs) composed of $M$ and $N$ antenna elements, respectively, following the FIM-based MIMO modeling perspective in~\cite{An_2025_FIM}. Unlike conventional rigid arrays, FIMs enable continuous morphological reconfiguration, in which the displacement of each antenna element is described by the morphing vectors $\mathbf{y}^{(T)} = [y_1^{(T)},y_2^{(T)}, \dots, y_M^{(T)}]$ and $\mathbf{y}^{(R)} = [y_1^{(R)},y_2^{(R)}, \dots, y_N^{(R)}]$.
We adopt a three-dimensional geometric channel model that characterizes signal propagation through $L$ distinct paths. Since the FIM can be arbitrarily oriented in three-dimensional space, a local coordinate frame must be established to consistently describe the position of each radiating element. To this end, we introduce an orthonormal basis $\{\mathbf{i}_t, \mathbf{j}_t, \mathbf{k}_t\}$ attached to the transmitting FIM, where $\mathbf{k}_t \in \mathbb{R}^3$ defines the surface normal along which morphing displacements occur, and $\mathbf{i}_t, \mathbf{j}_t \in \mathbb{R}^3$ span the array aperture plane. Furthermore, let $\phi_t \in [0, 2\pi)$ and $\theta_t \in [0, \pi]$ denote the azimuth and elevation angles associated with $\mathbf{k}_t$, respectively, both defined with respect to the transmitting FIM. Hence, $\mathbf{k}_t$ can be expressed as $\mathbf{k}_t =
[\sin\theta_t \cos\phi_t,
\sin\theta_t \sin\phi_t ,
\cos\theta_t]^T$.
By introducing a rotation angle $\varrho_t \in [0, 2\pi)$, which denotes the spin angle of the transmitting FIM and characterizes the rotation around its normal direction $\mathbf{k}_t$, the basis vectors $\mathbf{i}_t$ and $\mathbf{j}_t$ are written as
\begin{equation}\label{eq:i_t}
\mathbf{i}_t =
\begin{bmatrix}
\cos\theta_t \cos\phi_t \cos\varrho_t - \sin\phi_t \sin\varrho_t ,
\cos\theta_t \\ \times \sin\phi_t \cos\varrho_t + \cos\phi_t \sin\varrho_t ,
-\sin\theta_t \cos\varrho_t
\end{bmatrix}^T,
\end{equation}
\begin{equation}\label{eq:j_t}
\mathbf{j}_t =
\begin{bmatrix}
-\cos\theta_t \cos\phi_t \sin\varrho_t - \sin\phi_t \cos\varrho_t,
-\cos\theta_t \\ \times \sin\phi_t \sin\varrho_t + \cos\phi_t \cos\varrho_t ,
\sin\theta_t \sin\varrho_t
\end{bmatrix}^T.
\end{equation}

Let $M = M_x M_z$ be the total number of transmitting antennas, with $M_x$ and $M_z$ elements along $\mathbf{i}_t$ and $\mathbf{j}_t$. Taking $\mathbf{q}_1$ as reference, the $m$-th element lies at $\mathbf{q}_m = \mathbf{q}_1 + x_m^t \mathbf{i}_t + z_m^t \mathbf{j}_t$, where $x_m^t = d_{t,x}\,\mathrm{mod}(m{-}1, M_x)$, $z_m^t = d_{t,z}\,\lfloor (m{-}1)/M_x\rfloor$, and $d_{t,x},d_{t,z}$ are the inter-element spacings along $\mathbf{i}_t,\mathbf{j}_t$. Each element can be independently displaced along $\mathbf{k}_t$, giving $\tilde{\mathbf{q}}_m = \mathbf{q}_m + y_m^{(T)} \mathbf{k}_t$, with $-\tilde{y} \leq y_m^{(T)} \leq \tilde{y}$ and $\tilde{y} \geq 0$ the maximum deformation range.

Analogously, the receiving FIM is described following the same procedure adopted for the transmitting FIM in~\eqref{eq:i_t}--\eqref{eq:j_t}, with the orthonormal basis $\{\mathbf{i}_r, \mathbf{j}_r, \mathbf{k}_r\}$, azimuth and elevation angles $\phi_r \in [0, 2\pi)$, $\theta_r \in [0, \pi]$, and spin angle $\varrho_r \in [0, 2\pi)$ characterizing the receiving aperture.

Analogously, with $N = N_x N_z$ receiving antennas, the $n$-th element lies at $\mathbf{p}_n = \mathbf{p}_1 + x_n^r \mathbf{i}_r + z_n^r \mathbf{j}_r$, where $x_n^r = d_{r,x}\,\mathrm{mod}(n{-}1, N_x)$, $z_n^r = d_{r,z}\,\lfloor (n{-}1)/N_x\rfloor$, and $d_{r,x},d_{r,z}$ are the receiver inter-element spacings. Each element can be displaced along $\mathbf{k}_r$, giving $\tilde{\mathbf{p}}_n = \mathbf{p}_n + y_n^{(R)} \mathbf{k}_r$ with $-\tilde{y} \leq y_n^{(R)} \leq \tilde{y}$.

\subsection{Channel Model}

\begin{figure}[!t]
    \centering
    \includegraphics[width=.55\columnwidth]{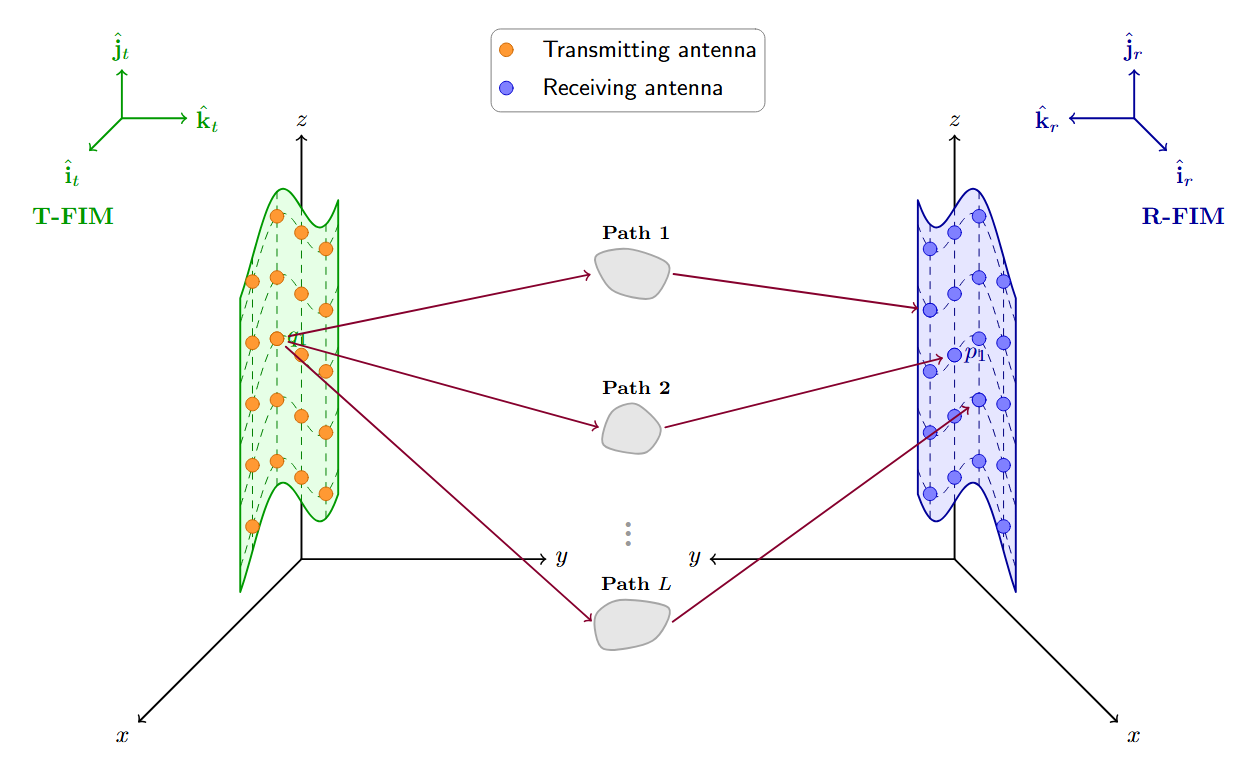}
    \caption{Schematic of a point-to-point MIMO system with FIM-based Tx and Rx.}
    \label{fig:system_representation}
\end{figure}

Let $\mathbf{H}(\mathbf{y}^{(T)},\mathbf{y}^{(R)}) \in \mathbb{C}^{N \times M}$ denote the FIM-assisted MIMO channel matrix between the transmitting and receiving FIMs. Unlike conventional rigid MIMO systems, the channel in FIM-assisted architectures depends on the three-dimensional surface configurations of both the transmitting and receiving FIMs.
The propagation direction associated with the $l$-th path of the transmitting FIM is given by $\mathbf{o}^t_l = [\sin \theta^t_l \cos\phi^t_l,\; \sin \theta^t_l \sin\phi^t_l,\; \cos \theta^t_l]^T$, for $l = 1,2,\dots,L$. The steering vector $\mathbf{b}(\phi_l^t,\theta_l^t) \in \mathbb{C}^{M}$ of the unmorphed transmitting FIM is then given by
\begin{equation}\label{eq:bt}
\begin{aligned}
\mathbf{b} \left( \phi_l^t, \theta_l^t \right) = &\left[ 1, \dots, e^{j\kappa \left( x_m^t \langle \mathbf{i}_t, \mathbf{o}_l^t \rangle + z_m^t \langle \mathbf{j}_t, \mathbf{o}_l^t \rangle \right)}, \right. \\
&\left. \dots, e^{j\kappa \left( x_M^t \langle \mathbf{i}_t, \mathbf{o}_l^t \rangle + z_M^t \langle \mathbf{j}_t, \mathbf{o}_l^t \rangle \right)} \right]^T,
\end{aligned}
\end{equation}
where $\kappa = 2\pi/\lambda$ denotes the wavenumber, with $\lambda$ being the carrier wavelength. Furthermore, the transmitting FIM can introduce additional three-dimensional deformation, allowing the position of each antenna element to be adjusted along its normal direction. In particular, the multiplicative response associated with the morphing configuration $\mathbf{y}^{(T)}$ is given by
\begin{equation}\label{eq:f(t)}
\mathbf{f}_t(\mathbf{y}^{(T)},\phi_l^t,\theta_l^t) =
\left[ 1, \dots, e^{j\kappa y_m^{(T)} \langle \mathbf{k}_t, \mathbf{o}_l^t \rangle}, \dots, e^{j\kappa y_M^{(T)} \langle \mathbf{k}_t, \mathbf{o}_l^t \rangle} \right]^T
\end{equation}
This multiplicative term captures the phase variation induced by the deformation of the transmitting FIM and modifies the original array response accordingly. Hence, the resulting steering vector of the morphed transmitting FIM is given as
\begin{equation}\label{eq:bt_tilde}
\tilde{\mathbf{b}} \left( \mathbf{y}^{(T)}, \phi_l^t, \theta_l^t \right)
=
\mathbf{b} \left( \phi_l^t, \theta_l^t \right)
\odot
\mathbf{f}_t \left( \mathbf{y}^{(T)}, \phi_l^t, \theta_l^t \right)
\end{equation}
for $l=1,2,\dots,L$.

Similarly, let $\phi_r \in [0, 2\pi)$ and $\theta_r \in [0, \pi]$ denote the azimuth and elevation angles associated with the receiving FIM. The propagation direction associated with the $l$-th path of the receiving FIM is given by $\mathbf{o}^r_l = [\sin \theta^r_l \cos\phi^r_l,\; \sin \theta^r_l \sin\phi^r_l,\; \cos \theta^r_l]^T$, for $l = 1,2,\dots,L$. The array steering vector $\mathbf{a}(\phi_l^r,\theta_l^r) \in \mathbb{C}^{N}$ of the unmorphed receiving FIM is then given by

\begin{equation}\label{eq:ar}
\begin{aligned}
\mathbf{a} \left( \phi_l^r, \theta_l^r \right) = &\left[ 1, \dots, e^{j\kappa \left( x_n^r \langle \mathbf{i}_r, \mathbf{o}_l^r \rangle + z_n^r \langle \mathbf{j}_r, \mathbf{o}_l^r \rangle \right)}, \right. \\
&\left. \dots, e^{j\kappa \left( x_N^r \langle \mathbf{i}_r, \mathbf{o}_l^r \rangle + z_N^r \langle \mathbf{j}_r, \mathbf{o}_l^r \rangle \right)} \right]^T
\end{aligned}
\end{equation}
Furthermore, the receiving FIM can also introduce additional three-dimensional deformation, enabling independent displacement of each antenna element along its normal direction. In particular, the corresponding multiplicative response associated with the morphing configuration $\mathbf{y}^{(R)}$ is given by
\begin{equation}\label{eq:f_r}
\mathbf{f}_r(\mathbf{y}^{(R)},\phi_l^r,\theta_l^r) =
\left[ 1, \dots, e^{j\kappa y_n^{(R)} \langle \mathbf{k}_r, \mathbf{o}_l^r \rangle}, \dots, e^{j\kappa y_N^{(R)} \langle \mathbf{k}_r, \mathbf{o}_l^r \rangle} \right]^T
\end{equation}
This multiplicative term captures the phase variation induced by the deformation of the receiving FIM and modifies its array response accordingly. Therefore, the resulting steering vector of the morphed receiving FIM is expressed as
\begin{equation}\label{eq:ar_tilde}
\tilde{\mathbf{a}} \left( \mathbf{y}^{(R)}, \phi_l^r, \theta_l^r \right)
=
\mathbf{a} \left( \phi_l^r, \theta_l^r \right)
\odot
\mathbf{f}_r \left( \mathbf{y}^{(R)}, \phi_l^r, \theta_l^r \right)
\end{equation}
For a given pair of morphing vectors $\{\mathbf{y}^{(R)}, \mathbf{y}^{(T)}\}$, the channel can be expressed as the superposition of $L$ propagation paths
\begin{equation}\label{eq:channel_model}
\mathbf{H}(\mathbf{y}^{(T)}, \mathbf{y}^{(R)}) =
\sum_{l=1}^{L} \alpha_l\,
\tilde{\mathbf{a}} \left( \mathbf{y}^{(R)}, \phi_l^{r}, \theta_l^{r} \right)
\tilde{\mathbf{b}}^{\mathrm{T}} \left( \mathbf{y}^{(T)}, \phi_l^{t}, \theta_l^{t} \right)
\end{equation}
where $\alpha_l$ denotes the complex gain of the $l$-th propagation path, for $l=1,\dots,L$. This geometric multipath representation, with $L \ll \min(M,N)$ dominant paths each parameterized by their angles of departure and arrival, is particularly well-suited to millimeter-wave (mmWave) communications, where the strong free-space attenuation and limited number of reflectors yield a low-rank channel naturally captured by a finite sum of rank-one steering-vector outer products.

\section{Training Protocol and Tensor Construction}
We adopt a split single time-scale training scheme in which $I+J$ pilot transmissions are split into two phases, sequentially decoupling the Tx and Rx morphing configurations and yielding $I+J$ third-order tensors. In the first $I$ slots, the Rx FIM morphs while the Tx FIM is fixed. With $\mathbf{S}=\mathrm{diag}(s_1,\dots,s_M)\in\mathbb{C}^{M\times M}$ the diagonal pilot matrix, the received signal at the $i$-th slot is
\begin{equation}\label{eq:Y1_signal}
\mathbf{Y}_i^{(1)} = \mathbf{D}_i(\mathbf{W})\,\mathbf{H}(\mathbf{y}_i^{(R)},\mathbf{y}_0^{(T)})\,\mathbf{S},
\end{equation}
where $\mathbf{Y}_i^{(1)}\in\mathbb{C}^{N\times M}$ and $\mathbf{W} \in \mathbb{C}^{N \times L}$ is the receive-side morphing response matrix whose $i$-th diagonal extraction $\mathbf{D}_i(\mathbf{W})$ models the phase variations of the $i$-th Rx morphing configuration. The additive noise term is omitted in the derivations; its impact on estimation accuracy is assessed in Section~V. Exploiting the UPA geometry, the array response is decomposed along the $x$ and $z$ dimensions, yielding the factorized representation
\begin{equation}\label{eq:Y1_factorized}
\mathbf{Y}_i^{(1)} =
\left[
\underbrace{\mathbf{D}_i(\mathbf{W}^{(x)})\tilde{\mathbf{A}}^{(x)}}_{\bar{\mathbf{A}}_i^{(x)}}
\diamond
\underbrace{\mathbf{D}_i(\mathbf{W}^{(z)})\tilde{\mathbf{A}}^{(z)}}_{\bar{\mathbf{A}}_i^{(z)}}
\right]
\mathbf{D}_{\alpha}
\mathbf{B}^{\mathrm{T}}
\mathbf{S}.
\end{equation}

After applying the matched filter $\mathbf{S}^{\mathrm{H}}$, the observation matrix $\mathbf{X}_i^{(1)}\in\mathbb{C}^{N\times M}$ is rearranged into a third-order tensor $\mathcal{X}_i^{(1)}\in\mathbb{C}^{N_x\times N_z\times M}$. The separable structure of the UPA and its morphing capability naturally lead to a low-rank tensor model. This justifies the use of the PARAFAC decomposition, which enables the joint estimation of channel parameters by exploiting its multi-way structure. Each tensor admits the PARAFAC decomposition
\begin{equation}\label{eq:tensor_phase1}
\mathcal{X}_i^{(1)} =
\mathcal{D}_{3,L}
\times_1 \bar{\mathbf{A}}_i^{(x)}
\times_2 \bar{\mathbf{A}}_i^{(z)}
\times_3 \mathbf{B},
\end{equation}
where $\mathcal{D}_{3,L}$ is a third-order superdiagonal tensor whose nonzero entries contain the complex path gains associated with the $L$ propagation paths.

In the second phase, the transmitter FIM varies its morphing configuration across $J$ time slots while the receiver remains fixed. The received signal matrix becomes
\begin{equation}\label{eq:Y2_signal}
\mathbf{Y}_j^{(2)} =
\mathbf{H}(\mathbf{y}_0^{(R)},\mathbf{y}_j^{(T)})
\mathbf{D}_j(\mathbf{F})
\mathbf{S}.
\end{equation}
where $\mathbf{F} \in \mathbb{C}^{M \times L}$ denotes the transmit-side morphing response matrix, whose $j$-th diagonal extraction through $\mathbf{D}_j(\mathbf{F})$ models the phase variations induced by the $j$-th morphing configuration of the transmitting FIM.
Exploiting the separability of the UPA geometry, the received signal can be written as
\begin{equation}\label{eq:Y2_factorized}
\mathbf{Y}_j^{(2)} =
\mathbf{A}
\mathbf{D}_{\alpha}
\left[
\underbrace{\mathbf{D}_j(\mathbf{F}^{(x)})\tilde{\mathbf{B}}^{(x)}}_{\bar{\mathbf{B}}_j^{(x)}}
\diamond
\underbrace{\mathbf{D}_j(\mathbf{F}^{(z)})\tilde{\mathbf{B}}^{(z)}}_{\bar{\mathbf{B}}_j^{(z)}}
\right]^{\mathrm{T}}
\mathbf{S}.
\end{equation}
After matched filtering, $\mathbf{X}_j^{(2)}$ is rearranged into a third-order tensor $\mathcal{X}_j^{(2)}\in\mathbb{C}^{N\times M_x\times M_z}$ with PARAFAC representation
\begin{equation}\label{eq:tensor_phase2}
\mathcal{X}_j^{(2)} =
\mathcal{D}_{3,L}
\times_1 \mathbf{A}
\times_2 \bar{\mathbf{B}}_j^{(x)}
\times_3 \bar{\mathbf{B}}_j^{(z)}.
\end{equation}

\section{Tensor-Based Channel Parameter Estimation}
Building on the tensor models of the previous section, we propose a two-phase PARAFAC Alternating Least Squares (ALS) algorithm~\cite{CLdA09} that sequentially estimates the morphing configurations and the static steering matrices by minimizing the Frobenius norm of the residual errors.

\subsection{Phase 1: Estimation of Receiver Components and $\mathbf{B}$}
In the first phase, the transmitter is fixed while the receiver FIM morphs over $I$ slots. The component matrices $\bar{\mathbf{A}}_i^{(x)}$ and $\bar{\mathbf{A}}_i^{(z)}$ are estimated per slot and the common transmit steering matrix $\mathbf{B}$ is then recovered by minimizing
\begin{equation}\label{eq:phase1_cost}
    \{ \hat{\bar{\mathbf{A}}}_i^{(x)}, \hat{\bar{\mathbf{A}}}_i^{(z)} \} = \underset{\bar{\mathbf{A}}_i^{(x)}, \bar{\mathbf{A}}_i^{(z)}}{\arg\min} \left\| \mathcal{X}_i^{(1)} - \text{PARAFAC}(\bar{\mathbf{A}}_i^{(x)}, \bar{\mathbf{A}}_i^{(z)}, \mathbf{B}) \right\|_{\mathrm{F}}^2,\nonumber
\end{equation}
for $i=1,2,\dots,I$. To solve this optimization problem, we adopt an ALS approach that exploits the three $n$-mode unfoldings of $\mathcal{X}_i^{(1)}$:

\begin{subequations}\label{eq:phase1_unfoldings}
\begin{align}
[\mathcal{X}_i^{(1)}]_{(1)} &= \bar{\mathbf{A}}_i^{(x)} (\bar{\mathbf{A}}_i^{(z)} \diamond \mathbf{B})^{\mathrm{T}}, \label{eq:phase1_unfolding1}\\
[\mathcal{X}_i^{(1)}]_{(2)} &= \bar{\mathbf{A}}_i^{(z)} (\bar{\mathbf{A}}_i^{(x)} \diamond \mathbf{B})^{\mathrm{T}}, \label{eq:phase1_unfolding2}\\
[\mathcal{X}_i^{(1)}]_{(3)} &= \mathbf{B} (\bar{\mathbf{A}}_i^{(x)} \diamond \bar{\mathbf{A}}_i^{(z)})^{\mathrm{T}}. \label{eq:phase1_unfolding3}
\end{align}
\end{subequations}
Assuming distinct path angles so the factor matrices are full column-rank, the PARAFAC decomposition is essentially unique (up to scaling/permutation) under the Kruskal-type condition $\mathrm{rank}(\bar{\mathbf{A}}_i^{(x)}) + \mathrm{rank}(\bar{\mathbf{A}}_i^{(z)}) + \mathrm{rank}(\mathbf{B}) \geq 2L + 2$, with ranks $\min(N_x,L)$, $\min(N_z,L)$, and $\min(M,L)$, respectively.

The ALS algorithm exploits~\eqref{eq:phase1_unfoldings} to update one factor matrix at a time in closed form. Starting from initial estimates of $\mathbf{B}$ and $\bar{\mathbf{A}}_i^{(z)}$, the receiver components are alternately updated by solving~\eqref{eq:phase1_unfolding1} and~\eqref{eq:phase1_unfolding2} in the least-squares sense:
\begin{align}
\hat{\bar{\mathbf{A}}}_i^{(x)} &= [\mathcal{X}_i^{(1)}]_{(1)} \left[(\bar{\mathbf{A}}_i^{(z)} \diamond \mathbf{B})^{\mathrm{T}}\right]^{\dagger}, \label{eq:Ax_update}\\
\hat{\bar{\mathbf{A}}}_i^{(z)} &= [\mathcal{X}_i^{(1)}]_{(2)} \left[(\bar{\mathbf{A}}_i^{(x)} \diamond \mathbf{B})^{\mathrm{T}}\right]^{\dagger}, \label{eq:Az_update}
\end{align}
for $i=1,\dots,I$, where $(\cdot)^{\dagger}$ denotes the Moore--Penrose pseudo-inverse.

Concatenating the third-mode unfoldings across the $I$ slots, we define $\boldsymbol{Y}_i^{(1)} = [\mathcal{X}_i^{(1)}]_{(3)}$ and  $\mathcal{Y}^{(1)} = \boldsymbol{Y}_1^{(1)} \sqcup_3 \cdots \sqcup_3 \boldsymbol{Y}_I^{(1)}$,
where $\mathcal{Y}^{(1)}$ has dimensions $M \times N_x N_z \times I$. The aggregated matrix $[\mathcal{Y}^{(1)}]_{(3)} = [\,[\mathcal{X}_1^{(1)}]_{(3)}\;\cdots\;[\mathcal{X}_I^{(1)}]_{(3)}\,]$ can then be factored as
\begin{equation}\label{eq:Y1_factorization_common}
    [\mathcal{Y}^{(1)}]_{(3)} = \mathbf{B} \underbrace{ \left[ (\bar{\mathbf{A}}_1^{(x)} \diamond \bar{\mathbf{A}}_1^{(z)})^{\mathrm{T}} \;\; \dots \;\; (\bar{\mathbf{A}}_I^{(x)} \diamond \bar{\mathbf{A}}_I^{(z)})^{\mathrm{T}} \right] }_{\bar{\mathbf{A}}},
\end{equation}
so that $\mathbf{B}$ is updated in closed form by the aggregated least-squares solution
\begin{equation}\label{eq:B_update}
    \mathbf{B} = [\mathcal{Y}^{(1)}]_{(3)} \bar{\mathbf{A}}^{\dagger}.
\end{equation}

\subsection{Phase 2: Estimation of Transmitter Components and $\mathbf{A}$}
In the second phase, the receiver is fixed while the transmitter FIM morphs over $J$ slots. Analogously, the component matrices $\bar{\mathbf{B}}_j^{(x)}$ and $\bar{\mathbf{B}}_j^{(z)}$ are estimated per slot and the common receive steering matrix $\mathbf{A}$ is then recovered by minimizing
\begin{equation}\label{eq:phase2_cost}
    \{ \hat{\bar{\mathbf{B}}}_j^{(x)}, \hat{\bar{\mathbf{B}}}_j^{(z)} \} = \underset{\bar{\mathbf{B}}_j^{(x)}, \bar{\mathbf{B}}_j^{(z)}}{\arg\min} \left\| \mathcal{X}_j^{(2)} - \text{PARAFAC}(\mathbf{A}, \bar{\mathbf{B}}_j^{(x)}, \bar{\mathbf{B}}_j^{(z)}) \right\|_{\mathrm{F}}^2,\nonumber
\end{equation}
for $j=1,2,\dots,J$.
To solve this optimization problem, we again apply the ALS method, exploiting the three $n$-mode unfoldings of $\mathcal{X}_j^{(2)}$:
\begin{subequations}\label{eq:phase2_unfoldings}
\begin{align}
[\mathcal{X}_j^{(2)}]_{(1)} &= \mathbf{A} (\bar{\mathbf{B}}_j^{(x)} \diamond \bar{\mathbf{B}}_j^{(z)})^{\mathrm{T}}, \label{eq:phase2_unfolding1}\\
[\mathcal{X}_j^{(2)}]_{(2)} &= \bar{\mathbf{B}}_j^{(x)} (\mathbf{A} \diamond \bar{\mathbf{B}}_j^{(z)})^{\mathrm{T}}, \label{eq:phase2_unfolding2}\\
[\mathcal{X}_j^{(2)}]_{(3)} &= \bar{\mathbf{B}}_j^{(z)} (\mathbf{A} \diamond \bar{\mathbf{B}}_j^{(x)})^{\mathrm{T}}. \label{eq:phase2_unfolding3}
\end{align}
\end{subequations}
An analogous Kruskal-type uniqueness condition holds for this phase, namely $\mathrm{rank}(\mathbf{A}) + \mathrm{rank}(\bar{\mathbf{B}}_j^{(x)}) + \mathrm{rank}(\bar{\mathbf{B}}_j^{(z)}) \geq 2L + 2$, with ranks $\min(N,L)$, $\min(M_x,L)$, and $\min(M_z,L)$, respectively.

Starting from initial estimates of $\mathbf{A}$ and $\bar{\mathbf{B}}_j^{(z)}$, the transmitter components are alternately updated by solving~\eqref{eq:phase2_unfolding2} and~\eqref{eq:phase2_unfolding3} in the least-squares sense:
\begin{align}
\hat{\bar{\mathbf{B}}}_j^{(x)} &= [\mathcal{X}_j^{(2)}]_{(2)} \left[(\mathbf{A} \diamond \bar{\mathbf{B}}_j^{(z)})^{\mathrm{T}}\right]^{\dagger}, \label{eq:Bx_update}\\
\hat{\bar{\mathbf{B}}}_j^{(z)} &= [\mathcal{X}_j^{(2)}]_{(3)} \left[(\mathbf{A} \diamond \bar{\mathbf{B}}_j^{(x)})^{\mathrm{T}}\right]^{\dagger}, \label{eq:Bz_update}
\end{align}
for $j=1,\dots,J$. Concatenating the first-mode unfoldings across the $J$ slots, we define $\boldsymbol{Y}_j^{(2)} = [\mathcal{X}_j^{(2)}]_{(1)}$ and $\mathcal{Y}^{(2)} = \boldsymbol{Y}_1^{(2)} \sqcup_1 \cdots \sqcup_1 \boldsymbol{Y}_J^{(2)}$, where $\mathcal{Y}^{(2)}$ has dimensions $N \times M_x M_z \times J$. The aggregated matrix $[\mathcal{Y}^{(2)}]_{(1)} = [\,[\mathcal{X}_1^{(2)}]_{(1)}\;\cdots\;[\mathcal{X}_J^{(2)}]_{(1)}\,]$ can then be factored as
\begin{equation}\label{eq:Y2_factorization_common}
    [\mathcal{Y}^{(2)}]_{(1)} = \mathbf{A} \underbrace{ \left[ (\bar{\mathbf{B}}_1^{(x)} \diamond \bar{\mathbf{B}}_1^{(z)})^{\mathrm{T}} \;\; \dots \;\; (\bar{\mathbf{B}}_J^{(x)} \diamond \bar{\mathbf{B}}_J^{(z)})^{\mathrm{T}} \right] }_{\bar{\mathbf{B}}},
\end{equation}
so that $\mathbf{A}$ is updated in closed form by the aggregated least-squares solution
\begin{equation}\label{eq:A_update}
    \mathbf{A} = [\mathcal{Y}^{(2)}]_{(1)} \bar{\mathbf{B}}^{\dagger}.
\end{equation}
The ALS monotonically decreases the Frobenius residual at each substep, with per-iteration cost $\mathcal{O}\big((I+J)\,NML\big)$~\cite{Favier_EUSIPICO}.

\begin{algorithm}[!t]
\caption{Proposed Two-Phase PARAFAC-ALS Algorithm}
\begin{algorithmic}[1]
\STATE \textbf{Input:} $\{\mathcal{X}_i^{(1)}\}_{i=1}^I$, $\{\mathcal{X}_j^{(2)}\}_{j=1}^J$, $L$.
\STATE \textbf{Initialize:} $\mathbf{A}$ and $\mathbf{B}$.
\REPEAT
\STATE \textit{Phase 1 -- Receiver estimation and update of $\mathbf{B}$:}
\FOR{$i=1,\dots,I$}
\STATE $\hat{\bar{\mathbf{A}}}_i^{(x)} \leftarrow [\mathcal{X}_i^{(1)}]_{(1)} \left[(\bar{\mathbf{A}}_i^{(z)} \diamond \mathbf{B})^{\mathrm{T}}\right]^{\dagger}$ \hfill \eqref{eq:Ax_update}
\STATE $\hat{\bar{\mathbf{A}}}_i^{(z)} \leftarrow [\mathcal{X}_i^{(1)}]_{(2)} \left[(\bar{\mathbf{A}}_i^{(x)} \diamond \mathbf{B})^{\mathrm{T}}\right]^{\dagger}$ \hfill \eqref{eq:Az_update}
\ENDFOR
\STATE $\mathbf{B} \leftarrow [\mathcal{Y}^{(1)}]_{(3)}\,\bar{\mathbf{A}}^{\dagger}$ \hfill \eqref{eq:B_update}
\STATE \textit{Phase 2 -- Transmitter estimation and update of $\mathbf{A}$:}
\FOR{$j=1,\dots,J$}
\STATE $\hat{\bar{\mathbf{B}}}_j^{(x)} \leftarrow [\mathcal{X}_j^{(2)}]_{(2)} \left[(\mathbf{A} \diamond \bar{\mathbf{B}}_j^{(z)})^{\mathrm{T}}\right]^{\dagger}$ \hfill \eqref{eq:Bx_update}
\STATE $\hat{\bar{\mathbf{B}}}_j^{(z)} \leftarrow [\mathcal{X}_j^{(2)}]_{(3)} \left[(\mathbf{A} \diamond \bar{\mathbf{B}}_j^{(x)})^{\mathrm{T}}\right]^{\dagger}$ \hfill \eqref{eq:Bz_update}
\ENDFOR
\STATE $\mathbf{A} \leftarrow [\mathcal{Y}^{(2)}]_{(1)}\,\bar{\mathbf{B}}^{\dagger}$ \hfill \eqref{eq:A_update}
\UNTIL{the reconstruction error variation falls below a preset threshold or the maximum number of iterations is reached}
\STATE \textbf{Output:} $\hat{\mathbf{A}}$, $\hat{\mathbf{B}}$, $\{\hat{\bar{\mathbf{A}}}_i^{(x)}\}_{i=1}^I$, $\{\hat{\bar{\mathbf{A}}}_i^{(z)}\}_{i=1}^I$, $\{\hat{\bar{\mathbf{B}}}_j^{(x)}\}_{j=1}^J$, $\{\hat{\bar{\mathbf{B}}}_j^{(z)}\}_{j=1}^J$.
\end{algorithmic}
\end{algorithm}

\section{Simulation Results}
This section discusses the impact of the physical morphing configuration design using the PARAFAC-based channel estimator. The performance is evaluated through the normalized mean square error (NMSE) of the estimated receive and transmit steering matrices, $\mathbf{A}$ and $\mathbf{B}$. The simulation setup consists of Monte Carlo simulations averaged over 1000 runs. Unless otherwise stated, the system considers $L = 3$ propagation paths, $4 \times 4$ FIM elements at both transmitter and receiver, and training time slots set to $I = 10$ and $J = 10$. The default SNR is set to $10$~dB. The mechanical displacement values are uniformly distributed within the maximum range up to $\tilde{y} = 1.0\lambda$. These values are chosen to avoid excessive physical deformation (i.e., preserving structural integrity) while providing sufficient phase diversity for the FIM configuration during the channel estimation task. Furthermore, the orientations of the transmitting and receiving FIMs are fixed throughout the simulations. In particular, the transmitting FIM adopts the orientation parameters $(\theta_t,\phi_t,\varrho_t)=(\pi/4,\pi/3,\pi/6)$, while the receiving FIM employs $(\theta_r,\phi_r,\varrho_r)=(\pi/3,\pi/6,-\pi/4)$. The propagation directions associated with each path are randomly generated for every Monte Carlo realization, where the azimuth and elevation angles are independently drawn from uniform distributions over the interval $[0,\pi]$.

\begin{figure}[!t]
    \centering
    \includegraphics[width=.48\columnwidth]{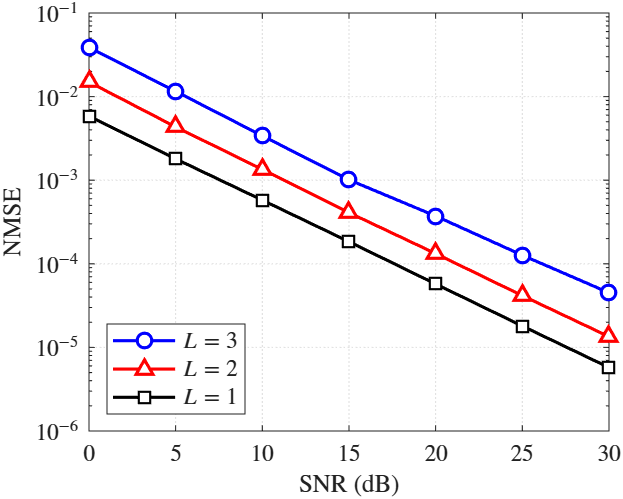}
    \caption{Impact of the number of propagation paths on the NMSE as a function of SNR.}
    \label{fig:snr_paths}
\end{figure}

\begin{figure}[!t]
    \centering
    \includegraphics[width=.48\columnwidth]{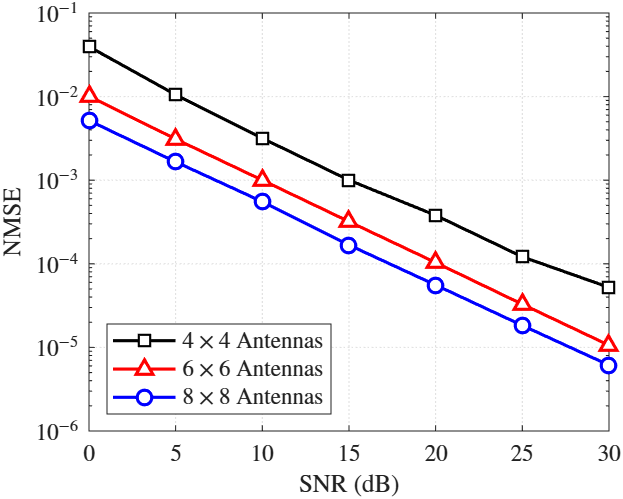}
    \caption{Impact of the number of antennas on the NMSE performance for different array configurations.}
    \label{fig:antennas_paths}
\end{figure}

\begin{figure}[!t]
    \centering
    \includegraphics[width=.48\columnwidth]{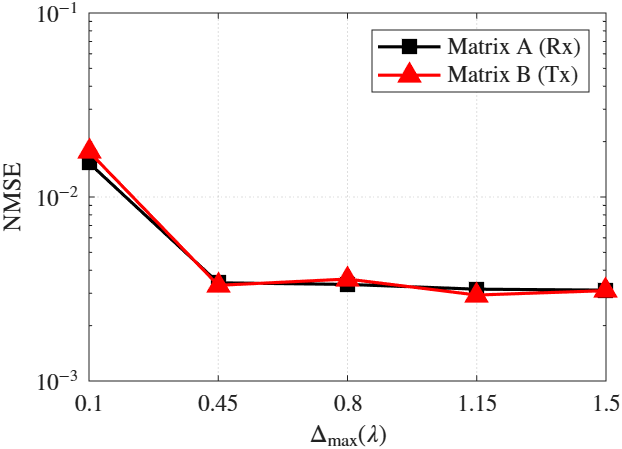}
    \caption{Impact of the morphing range on the NMSE for both receive and transmit matrices.}
    \label{fig:nmse_dual}
\end{figure}

Figures~\ref{fig:snr_paths} and~\ref{fig:antennas_paths} illustrate the channel estimation performance under different propagation conditions and array configurations. In Figure~\ref{fig:snr_paths}, the estimation accuracy decreases as the number of propagation paths ($L$) increases, since a larger number of multipath components makes the channel reconstruction more challenging. On the other hand, Figure~\ref{fig:antennas_paths} shows that increasing the number of antennas improves the estimation performance by providing higher spatial diversity and additional observations for the estimation process. Figure~\ref{fig:nmse_dual} presents the impact of the morphing range on the estimation accuracy. It is observed that small morphing ranges ($\tilde{y}$) lead to poor estimation performance due to the limited phase diversity introduced by the FIM configurations, which increases the correlation among the received observations. In contrast, larger morphing ranges improve the estimation accuracy because they provide more diverse channel observations. However, excessively large deformations may become impractical due to the physical limitations of the FIM structure. Overall, these results indicate that the morphing range emerges as a key design parameter and must be jointly tuned with the array dimension and the channel's multipath richness to achieve accurate CSI recovery.


\section{Conclusion}
We proposed a tensor-based channel estimation framework for point-to-point MIMO systems with active FIM arrays at both ends. A split single time-scale training protocol decouples the morphing configurations of the two FIMs and yields low-rank tensor models, from which a PARAFAC-ALS algorithm jointly estimates the morphed steering matrices and path gains while preserving the multidimensional structure induced by the FIM geometry. Numerical results show accurate CSI recovery across different system configurations; restrictive morphing ranges reduce phase diversity and degrade accuracy, while larger arrays compensate.
These observations expose a fundamental trade-off between mechanical flexibility, training overhead, and estimation performance in FIM-assisted MIMO systems.

\renewcommand\baselinestretch{.93}
\bibliographystyle{IEEEtran}
\bibliography{refs}

\end{document}